\documentstyle[12pt]{article}

\def\beq{\begin{equation}}
\def\eeq{\end{equation}}
\def\beqa{\begin{eqnarray}}
\def\eeqa{\end{eqnarray}} 

\begin{document}
 
\begin{titlepage}

\begin{center}

{\LARGE
Combining exclusive semi-leptonic and hadronic $B$ decays
to measure $|V_{ub}|$.}\\ [.5in]
{\large Jo\~{a}o M. Soares}\\ [.1in]
{\small
Department of Physics and Astronomy, University of Massachusetts,\\
Amherst, MA 01003-4525}\\ [.5in]

{\normalsize\bf
Abstract}\\ [.2in]

\end{center}

{\small
The Cabibbo-Kobayashi-Maskawa matrix element $|V_{ub}|$ can be extracted from 
the rate for the semi-leptonic decay $B \to \pi l^- \bar {\nu}_l$, with little 
theoretical uncertainty, provided the hadronic form factor for the $B \to \pi$
transition can be measured from some other $B$ decay. In here, we suggest 
using the decay $B \to \pi J/\psi$. This is a color suppressed decay, and it 
cannot be properly described within the usual factorization approximation; we 
use instead a simple and very general phenomenological model for the $b d 
J/\psi$ vertex. In order to relate the hadronic form factors in the $B \to \pi
J/\psi$ and $B \to \pi l^- \bar{\nu}_l$ decays, we use form factor relations 
that hold for heavy-to-light transitions at large recoil.}

\end{titlepage}

\section*{ }

The main difficulty in extracting the Cabibbo-Kobayashi-Maskawa (CKM) matrix 
element $|V_{ub}|$, from exclusive semi-leptonic decays such as $\bar{B}^0_d 
\to \pi^+ l^- \bar{\nu}_l$, is the theoretical uncertainty associated with the 
form factor $f_1(q^2)$, in the differential decay rate
\beq
\frac{d}{dq^2}\Gamma(\bar{B}^0_d \to \pi^+ l^- \bar{\nu}_l) 
= \frac{G_F^2}{24\pi^3} |V_{ub}|^2 |\vec{p}_\pi|^3 |f_1(q^2)|^2
\label{eq:1}
\eeq
($q \equiv p_B - p_\pi$, and $|\vec{p}_\pi|$ is the three-momentum of the pion
in the $B$ meson rest-frame). One way to overcome this problem is to compare
the semi-leptonic decay to some other $B$ decay, which involves the same 
hadronic transition, but is proportional to a different CKM matrix element
\cite{BurdmanDonoghue}. In here, we suggest using the tree level hadronic 
decay $B^- \to \pi^- J/\psi$. This decay has been observed recently, at both 
CLEO \cite{bpsipiCLEO} and CDF \cite{bpsipiCDF}; the present value of the 
branching ratio is $B(B^- \to \pi^- J/\psi) = (4.4 \pm 2.4) \times 10^{-5}$ 
\cite{PDG96}. The semi-leptonic decay $\bar{B}^0_d \to \pi^+ l^- \bar{\nu}_l$ 
has been seen at CLEO \cite{bpilnuCLEO}, with a preliminary branching ratio 
$B(\bar{B}^0_d \to \pi^+ l^- \bar{\nu}_l) \simeq 1.4 \times 10^{-4}$. More 
data will be necessary, in order to determine the differential decay rate in 
eq.~\ref{eq:1}.

The decays $B^- \to \pi^- J/\psi$ or $\rho^- J/\psi$, as their Cabibbo-allowed
analogues $B^- \to K^- J/\psi$ or $K^{\ast -} J/\psi$, and all other $B$ decays
to charmonium states, are color suppressed tree level decays. They are 
notorious for the failure of the common factorization procedure \cite
{factorization} to predict decay rates or polarization ratios \cite
{nonfactorization}. We will use instead the phenomenological model proposed in
ref.~\cite{bspsi} to describe these decays; it allows for corrections to the 
factorization result to be included in a simple and economical way. According 
to that model, and assuming no significant spectator effects, the $B$ decays 
to $J/\psi$ stem from an effective $b q J/\psi$ vertex ($q = s$ or $d$, for 
the Cabibbo-allowed and suppressed decays, respectively)
\beqa
\Lambda^\mu_{bqJ/\psi} &=& - \frac{G_F}{\sqrt{2}} \: V_{cb}V_{cq}^\ast \: 
(C_2 + \frac{1}{3} C_1)\:
[ g_0 \: q^\mu \not{q} (1 - \gamma_5)  \nonumber\\ 
& & + g_1 \: (m_{J/\psi}^2 g^{\mu\nu} - q^\mu q^\nu) \gamma_\nu  (1 - \gamma_5)
+  g_2 \: m_b i \sigma^{\mu\nu} q_\nu  (1 + \gamma_5)  ] ,
\label{eq:2}
\eeqa
where $C_{1,2}$ are the Wilson coefficients in the weak Hamiltonian, and $q = 
p_{J/\psi}$. This is the most general expression for the vertex, when, as we 
do in here, the mass of the light quark $q$ is neglected. In the factorization
approximation, $g_1 = g_0 = f_{J/\psi}/m_{J/\psi}$ and $g_2 = 0$; however, the
coefficients $g_{0,1,2}$ deviate from these values due to both perturbative 
and non-perturbative gluon exchanges. In particular, such QCD effects will 
generate the Lorentz structure associated with $g_2$; that is absent from the 
factorization result, but it is essential to fit the polarization data \cite
{bspsi}. In here, the coefficients $g_1$ and $g_2$ are to be determined 
empirically, at $q^2 = m_{J/\psi}^2$, from the data for the $B$ meson decays 
into $J/\psi$. The term proportional to the form factor $g_0$ does not 
contribute to the decay amplitudes, and so $g_0$ will be left undetermined.

As long as final state interactions do not play a significant role \cite{FSI},
it follows from eq.~\ref{eq:2} that the amplitude for the $B^- \to \pi^- 
J/\psi$ decay is given by
\beqa
A(B^- \rightarrow \pi^- J/\psi) &=&
- \frac{G_F}{\sqrt{2}} V_{cb} V_{cd}^\ast (C_2 + \frac{1}{3} C_1) 
2 m_B |\vec{p}_\pi| m_{J/\psi} \nonumber\\
& & \times \left[ g_1 f_1(m_{J/\psi}^2) +  
g_2 m_b s(m_{J/\psi}^2) \right] 
\label{eq:3}
\eeqa
($|\vec{p}_\pi|$ is the three-momentum of the pion in the B meson rest-frame);
$f_1(q^2)$ is the same form factor as in the semi-leptonic decay $\bar{B}^0_d 
\to \pi^+ l^- \bar{\nu}_l$, and $s(q^2)$ is the form factor associated with 
the Lorentz structure of the $g_2$  term in eq.~\ref{eq:2}. In the $m_d \to 0$
limit that we consider in here, these form factors satisfy the relation
\beq
f_1(q^2) = - (m_B-E_\pi+|\vec{p}_\pi|) s(q^2) \ ,
\label{eq:4}
\eeq
that was derived in ref.~\cite{heavytolight} (but see also the earlier work 
by Stech, in ref.~\cite{Stech}), from the constituent quark picture for the 
hadronic transition. This, and other form factor relations for heavy-to-light 
transitions, follow in the static limit for the heavy $b$ quark and the 
ultra-relativistic limit for the recoiling light quark; they are independent 
of the exact form of the wave-functions of the mesons. They hold best at large 
recoil momenta, as is the case for the $B \to \pi$ transition at $q^2 = 
m_{J/\psi}^2$ \cite{heavytolight}. The $B^- \to \pi^- J/\psi$ decay rate is 
then
\beqa
\Gamma(B^- \to \pi^- J/\psi) &=& 
\frac{G_F^2}{4 \pi} |V_{cb}|^2 \sin^2 \theta_c 
(C_2 + \frac{1}{3} C_1)^2
|\vec{p}_\pi|^3 m_{J/\psi}^2 \nonumber \\
& & \times \left| f_1(m_{J/\psi}^2) g_1
\left( 1 - \frac{g_2}{g_1} \right) \right|^2 
\label{eq:5}
\eeqa
(for simplicity, we ignore the small pion mass, but this is not necessary). 

This expression can now be used to eliminate from eq.~\ref{eq:1} the 
dependence on the hadronic matrix element $f_1$. We obtain
\beqa
\lefteqn{\frac{1}{\Gamma(B^- \to \pi^- J/\psi)} 
\left[ \frac{d}{dz} \Gamma(\bar{B}^0_d \to \pi^+ l^- \bar{\nu}_l)
\right]_{z = \frac{m_{J/\psi}^2}{m_B^2}}} \nonumber \\
& & = \frac{1}{6\pi^2} \left|\frac{V_{ub}}{V_{cb}}\right|^2 
\frac{1}{\sin^2 \theta_c}  \frac{m_B^2}{m_{J/\psi}^2}
\left| (C_2 + \frac{1}{3} C_1)
g_1 \left( 1 - \frac{g_2}{g_1} \right) \right|^{-2} \ ,
\label{eq:6}
\eeqa
where $z \equiv q^2/m_B^2$; this is our main result. The remaining task is to 
extract from the data for the $B$ decays to $J/\psi$ the value of the 
parameters $g_{1,2}$. This was done in ref.~\cite{bspsi}; however, the 
derivation in there relied on a specific model for the $q^2$ dependence of the
hadronic form factors. Once again, we can use the heavy-to-light form factor 
relations of refs.~\cite{heavytolight} and \cite{Stech}, and avoid the model 
dependence in the evaluation of $g_{1,2}$. 

The ratio $g_2/g_1$ can be extracted from the polarization in the decay $B \to 
K^\ast J/\psi$. Using both the effective $b s J/\psi$ vertex and the 
heavy-to-light form factor relations, we obtain for the $B \to K^\ast J/\psi$ 
helicity amplitudes
\beqa
\frac{A_+}{A_0} &=& 0 
\label{eq:7}  \\
\frac{A_-}{A_0} &=& -\frac{2 m_{J/\psi}}{m_B - E_{K^\ast} + 
|\vec{p}_{K^\ast}|} \left( 1 - \frac{g_2}{g_1}
\frac{m_B(m_B - E_{K^\ast} + |\vec{p}_{K^\ast}|)}{m_{J/\psi}^2} \right)
\nonumber \\
& & \times \left( 1 - \frac{g_2}{g_1} \frac{m_B}{m_B - E_{K^\ast} 
+ |\vec{p}_{K^\ast}|} \right)^{-1} \ ,
\label{eq:8}
\eeqa
where $E_{K^\ast}$ and $|\vec{p}_{K^\ast}|$ are the energy and three-momentum 
of the $K^\ast$ in the $B$ meson rest-frame. The polarization ratio
\beq
\frac{\Gamma_L}{\Gamma} \equiv \frac{|A_0|^2}{|A_0|^2+|A_-|^2+|A_+|^2}
= \frac{1}{1+|A_-/A_0|^2}
\label{eq:9}
\eeq
can then be used to determine $g_2/g_1$. With $\Gamma_L/\Gamma = 0.78 \pm 
0.07$ \cite{polarization}, we obtain $g_2/g_1 = 0.24 \pm 0.03$ (the two-fold
ambiguity in the solution is resolved using the value found in ref.~\cite
{bspsi}).

We should point out that, in general, the ratio $g_2/g_1$ may have a 
nontrivial phase \cite{phase} that we ignore, for now. It can be determined 
from a more detailed study of the angular correlations in the $B \to K^\ast 
J/\psi \to (K\pi)(e^+e^-)$ decay. The distribution in the angles $\theta_K$, 
$\theta_{e^+}$ and $\phi$ (respectively, the polar angles of the 
$K$ and $e^+$ momenta with respect to the momenta of the parent particles 
$K^\ast$ and $J/\psi$, and the azimuthal angle between the $K^\ast$ and 
$J/\psi$ decay planes) is determined by the quantities \cite{KramerPalmer}
\beqa
\alpha_1 &\equiv& \frac{Re(A_+ A_0^\ast + A_- A_0^\ast)}
{|A_0|^2 + |A_-|^2 + |A_+|^2}
\label{eq:10} \\
\beta_1 &\equiv& \frac{Im(A_+ A_0^\ast - A_- A_0^\ast)}
{|A_0|^2 + |A_-|^2 + |A_+|^2}
\label{eq:11} \\
\alpha_2 &\equiv& \frac{Re(A_+ A_-^\ast)}
{|A_0|^2 + |A_-|^2 + |A_+|^2}
\label{eq:12} \\
\beta_2 &\equiv& \frac{Im(A_+ A_-^\ast)}
{|A_0|^2 + |A_-|^2 + |A_+|^2}  \ .
\label{eq:13}
\eeqa
A measurement of the coefficients $\alpha_1$ and $\beta_1$, which in our model
are
\beqa
\alpha_1 &=& \frac{\Gamma_L}{\Gamma} Re \left( \frac{A_-}{A_0} \right)
\label{eq:14} \\
\beta_1 &=& - \frac{\Gamma_L}{\Gamma} Im \left( \frac{A_-}{A_0} \right) \ ,
\label{eq:15}
\eeqa
will allow a determination of the amplitude and phase of $g_2/g_1$. On the 
other hand, a measurement of the coefficients $\alpha_2$ and $\beta_2$, which 
vanish in our model, provides an interesting test of the approximation $m_s \to
0$, that was used in both the $b s J/\psi$ vertex and in the $B \to K^\ast$ 
form factors. 

As for $|g_1|$, it is determined from the branching ratio for the inclusive 
decay $B \to J/\psi + {\rm anything}$,
\beqa
\lefteqn{B(B \to J/\psi + {\rm anything}) = [\Gamma (b \to s J/\psi) + 
 \Gamma (b \to d J/\psi)]/\Gamma} \nonumber \\
&=& \frac{G_F^2}{16 \pi} \tau_B |V_{cb}|^2
(C_2 + \frac{1}{3} C_1)^2 m_b^5
\left( 1 - \frac{m_{J/\psi}^2}{m_b^2} \right)^2
\frac{m_{J/\psi}^2}{m_b^2} \nonumber \\
& & \times |g_1|^2
\left( |1 - \frac{g_2}{g_1}|^2 + 2 \frac{m_{J/\psi}^2}{m_b^2}
 |1 - \frac{g_2}{g_1}\frac{m_b^2}{m_{J/\psi}^2}|^2  \right) \ . 
\label{eq:16}
\eeqa
With $B(B \to J/\psi + {\rm anything}) = (0.82 \pm 0.08)\%$ \cite
{factorization}, the value that was found above for $g_2/g_1$, and taking $m_b
= m_B$ and $\tau_B = 1.6$ psec, we obtain $|V_{cb} (C_2 + C_1/3) g_1| = 
(1.81 \pm 0.14) \times 10^{-3}$.

Finally, the expression in eq.~\ref{eq:6} becomes
\beqa
\lefteqn{\frac{1}{\Gamma(B^- \to \pi^- J/\psi)}
\left[ \frac{d}{dz} \Gamma(\bar{B}^0_d \to \pi^+ l^- \bar{\nu}_l)
\right]_{z = \frac{m_{J/\psi}^2}{m_B^2}}}  \nonumber \\
& & = |V_{ub}|^2 \times (0.54 \pm 0.06) \times 10^6  \ ,
\label{eq:17}
\eeqa
which can then be used to determine $|V_{ub}|$ (the error corresponds to the
present experimental uncertainty in our determination of the parameters 
$g_{1,2}$). Corrections to this result are necessary, if the ratio $g_2/g_1$ 
proves to have a significant phase. The residual theoretical uncertainty in 
our result is that associated with the heavy-to-light form factor relations of 
refs.~\cite{heavytolight} and \cite{Stech}, which are valid in the limit of a
static heavy $b$ quark and a massless recoiling quark. From the analysis in 
ref.~\cite{heavytolight}, it is expected that this approximation holds well 
for the $B \to \pi$ transition, throughout most of the kinematic range and, in
particular, for  $q^2 = m_{J/\psi}^2$. Corrections to the form factor 
relations will be larger, in the case of the $B \to \rho$ or $B \to K^\ast$ 
transitions (hence our choice of the semileptonic decay $B \to \pi l^- \bar 
{\nu}_l$, rather than $B \to \rho l^- \bar{\nu}_l$). These transitions play a 
role in the evaluation of $g_2/g_1$, and so this should be the dominant source
of the theoretical uncertainty in our final result. As we pointed out before, 
the size of that uncertainty can be probed experimentally, with a detailed 
angular analysis of the $B \to K^\ast J/\psi$ or $B \to \rho J/\psi$ decays.

\section*{}

This work was supported in part by a grant from the National Science 
Foundation.


\begin{thebibliography}{Abcd:efgh}
\bibitem{FSI}
To this date, no final state interaction phases have been detected in $B$ 
decay amplitudes. This suggests that, contrary to the case of $K$ or $D$ 
mesons, final state interaction effects are small in $B$ decays (at least in 
the dominant tree level decays).
\bibitem{polarization}
This is the average of the ARGUS, CDF and CLEO results, as given in ref.~\cite
{factorization}.
\end{thebibliography}
\end{document}